# Computational Particle Physics for Event Generators and Data Analysis

**Denis Perret-Gallix**

LAPP (IN2P3/CNRS), France

E-mail: denis.perret-gallix@in2p3.fr

**Abstract**. High-energy physics data analysis relies heavily on the comparison between experimental and simulated data as stressed lately by the Higgs search at LHC and the recent identification of a Higgs-like new boson. The first link in the full simulation chain is the event generation both for background and for expected signals. Nowadays event generators are based on the automatic computation of matrix element or amplitude for each process of interest.

Moreover, recent analysis techniques based on the matrix element likelihood method assign probabilities for every event to belong to any of a given set of possible processes. This method originally used for the top mass measurement, although computing intensive, has shown its efficiency at LHC to extract the new boson signal from the background.

Serving both needs, the automatic calculation of matrix element is therefore more than ever of prime importance for particle physics. Initiated in the 80's, the techniques have matured for the lowest order calculations (tree-level), but become complex and CPU time consuming when higher order calculations involving loop diagrams are necessary like for QCD processes at LHC. New calculation techniques for next-to-leading order (NLO) have surfaced making possible the generation of processes with many final state particles (up to 6). If NLO calculations are in many cases under control, although not yet fully automatic, even higher precision calculations involving processes at 2-loops or more remain a big challenge.

After a short introduction to particle physics and to the related theoretical framework, we will review some of the computing techniques that have been developed to make these calculations automatic. The main available packages and some of the most important applications for simulation and data analysis, in particular at LHC will also be summarized (see CCP2012 slides [1]).

## 1. Particle Physics goals and means

Particle physics targets the study of the ultimate basic elements of matter and of the fundamental forces generated by or acting on them; the smaller the element is, the smaller the probe wavelength must be and, therefore, the larger the interaction energy. Actually, particle physics or more precisely its high-energy physics branch tends to reproduce on earth the range of energies that was prevalent when matter did start to form, at a very early time of the Universe history (fig. 1), namely some $10^{-10}$ s after the Big Bang. There is a profound duality between the today state of matter probed by high-energy colliders and the state of the Universe shortly after its creation, 13.7 billion years ago.

The highest man-made particle colliding energy reaches, today, 8 TeV (Tera Electron Volt ~ thousand times the proton mass) on the way to the nominal 14 TeV expected at the large hadron collider (LHC [2]) at CERN [3], Geneva (Switzerland). This energy, however, remains many orders of magnitude smaller than the typical interaction energy occurring in the Universe when quarks, gluons and electrons were moving in a confinement free "hot soup". From this even earlier Universe energy we only have a glimpse, today, through the very high energy cosmic rays (more than $10^8$ TeV) hitting the upper earth atmosphere. Producing huge particle showers they are detected by large area detectors on earth surface like at the 3000 km$^2$ Pierre Auger Observatory [4] or observed by satellite surveying the earth as proposed by the GEM-Euso Collaboration [5].

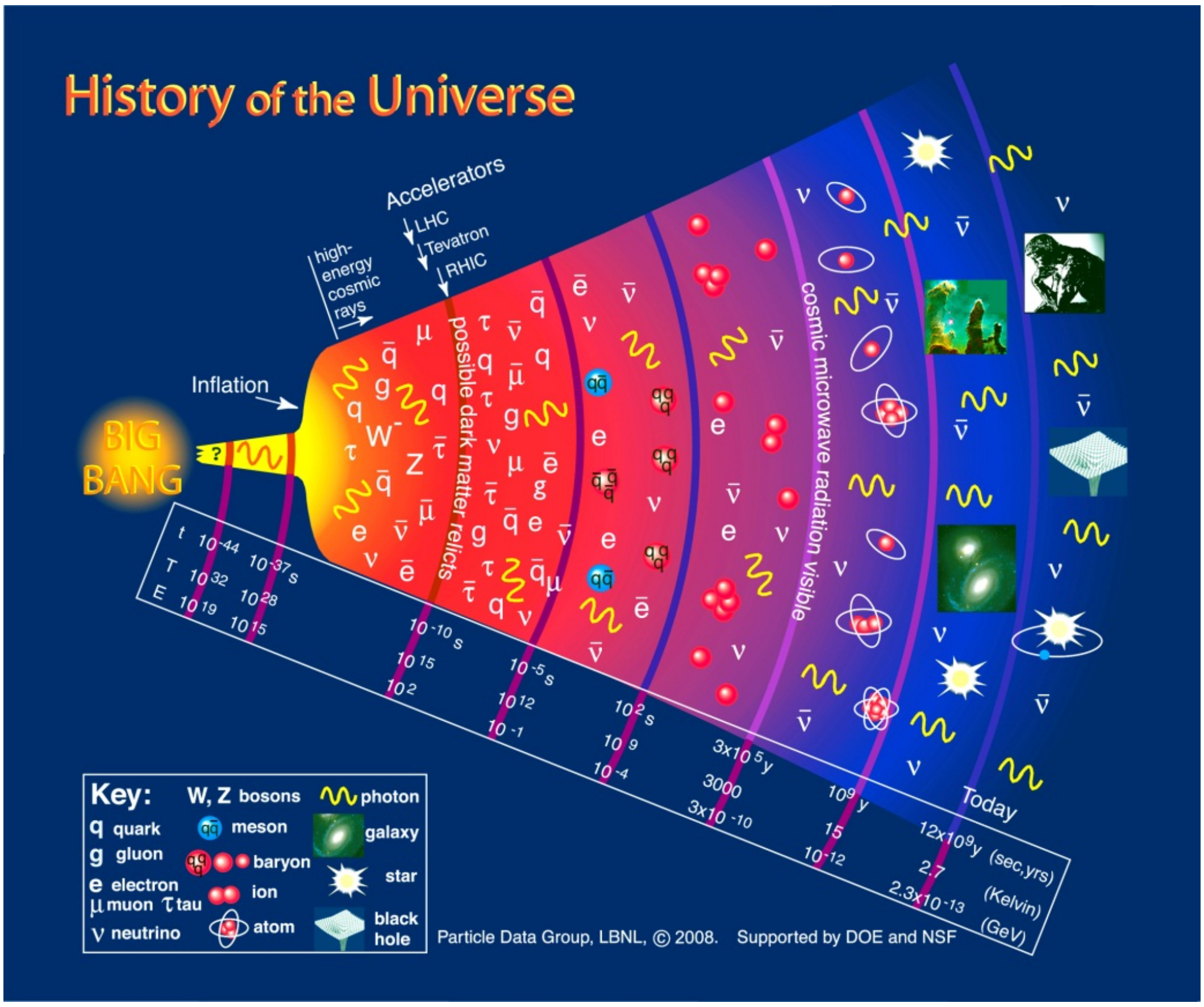

***Figure 1: History of the Universe***

Located at CERN, the LHC (Fig. 2) is the latest built proton-proton collider. It is housed in a 27 km circular, 100 m deep underground tunnel in the Geneva region. The LHC benefits from a large accelerator infrastructure progressively installed at CERN since it was founded back in the 50's. The proton bunches produced from a hydrogen source are accelerated by the Linac and Booster. Then, the PS and SPS increase the energy up to 450 GeV before entering the LHC. Four experiments are located in the LHC ring: two general purpose ATLAS [6] and CMS [7] and two dedicated LHCb [8] for the meson B physics and ALICE [9] for hadronic physics when protons are replaced by heavy ions (e.g. lead-lead collision) in the LHC.

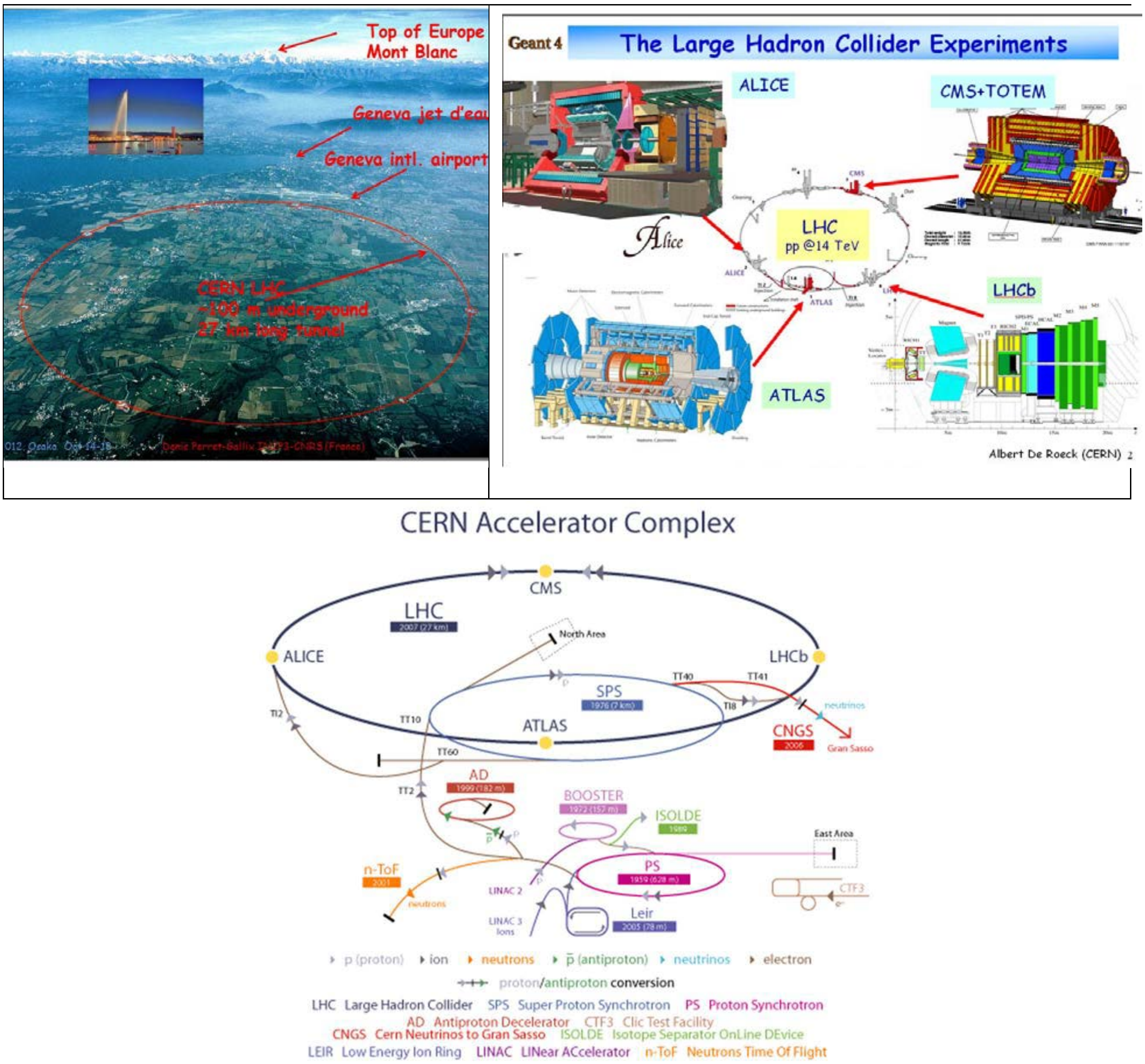

*Figure 2: LHC and experiments at CERN*

The particle physics overall picture is now embedded in a very solid theory that, shyly enough, we keep calling a Model. The Standard Model (SM) is maybe one of the most developed and rugged theory of any physics fields left alone other research domains. This is probably due to the fact that dealing with fundamental processes, generally, does not lead to the complexity of n-body problems generally intractable, although supercomputers are opening new hopes, at least for small n.

But that does not mean that the theory is simple when it comes to numerical evaluations. Fortunately methods have been developed based on the underlying quantum field theory to make precise computations. Until now, the experimental data have always either confirmed the theoretical predictions or brought new insight reinforcing the general view.

This success story seems to continue with the recent discovery of a "predicted" boson at the CERN LHC, last July (ATLAS [10], CMS [11]). Nobody yet dares calling it "Higgs", the last missing ingredient of the SM at current scales. We will soon learn from the experiments whether or not this new boson is the actual SM Higgs proposed as a consequence of a mechanism giving mass to elementary particles [12] [13] [14] [15].

This discovery is materialized by the small bump popping out of the recombined γγ mass histograms (fig.3) for both ATLAS [10] and CMS [11]. These simple plots summarize the extensive computational undertaking behind high-energy physics data analysis. For example, in the CMS experiment, the dots and the error bars represent the observed number of events in a given mass bin. The numbers come from a long chain of data analysis beginning with the trigger event selection, followed by the event reconstruction, the event filtering and analysis for the millions of events recorded by the experiments.

The yellow/green bands represent the best estimation of the so-called background events. They are based on a fit of the data points away from the bump. This fit is also compared to simulated events belonging to background processes. In red, the sum of background and simulated Higgs signal is shown. It is a clear demonstration of the interplay between experimental and simulated data.

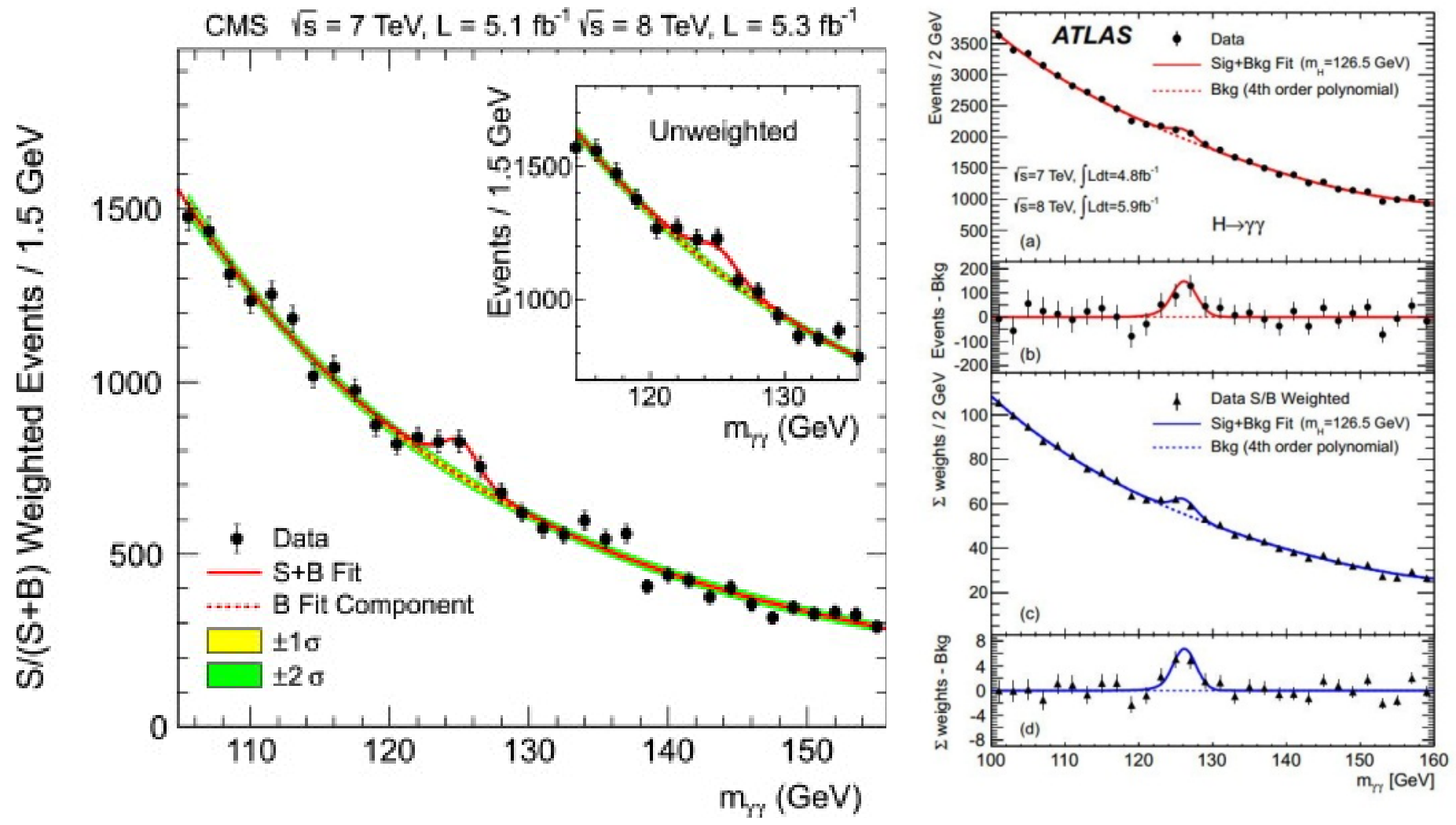

***Figure 3: Higgs-like boson signal from CMS and ATLAS (CERN)***

Most of the data simulation and analysis is performed on a distributed data grid system called the WLCG (Worldwide LHC Computing Grid) [16]. This infrastructure based on tiers computing centres has been installed all around the world to absorb the heavy computing load of the high LHC statistics. The tier 0 is located at the data production centre (CERN) and absorbs 20% of the load, 11 tiers 1 connected through direct high bandwidth fibre links (10 Gb/s) are scattered across the main participating countries/regions and ~ 140 tiers 2 are providing the specific computing support necessary to research teams. The WLCG provides more than 250,000 processors, and close to 150 PB of disk storage from over 150 sites in 34 countries, producing a massive distributed computing infrastructure that support the needs of more than 8,000 physicists.

In sect. 2, we will describe the general theoretical framework. Sect. 3 is a short presentation of the perturbative calculus and of the recent higher precision calculation breakthroughs. Then, in the following sections, like Russian dolls, we will go deeper in the experimental simulation chain (sect. 4): with the description the event formation phases (sect. 5), of the various steps leading to the building of event generator (sect. 6), including the automatic calculation of matrix elements (sect. 7) and their multi-dimensional integration giving the cross-sections (sect. 8). Sect. 9 describes the more recent usage of the same matrix element expressions for a more complete event analysis. Finally, sect. 10 lists some of the computational techniques that are used and shared with other research fields.

## 2. Perturbative and not perturbative? that is the … calculation.

Let us first replace this computational activity in the general framework of particle physics and its Standard Model. There are basically two domains of calculation depending on the actual value of the QCD[1] coupling constant $\alpha_s$. This constant is actually "running" from small values close to zero in the "asymptotic freedom" regime, namely at high energy when probing the deep inside of hadrons, to large values ($\alpha_s \sim 1$) in the "confinement" region where the partons (quarks and gluons) are kept inside the hadrons.

If $\alpha_s$ is small when, for example, 2 partons, almost free inside the protons, interact, the solutions of the Schrödinger equation which represents the time evolution of the quantum state can be approximated by a series expansion in power of $\alpha_s$. This is the so-called perturbative domain. Each term of the series corresponds to a level of precision, the higher the number of terms the more precise the calculation. We will see later also that each term of the series can be graphically represented by a set of "Feynman" diagrams [17], all having similar characteristics. This domain describes the dynamics of QCD interactions. This is the branch of computational particle physics that will be discussed in this paper.

Conversely, if $\alpha_s$ is large ~ 1, the hadron quantum system is probed at the confinement limit. Quarks and gluons are all strongly coupled and cannot escape the hadron boundary. Here the interest is more focused on the hadron as a whole. Global values like hadron masses are computed. The perturbative approximation does not apply though and the Schrödinger equation must be solved exactly. However, this has proved to be analytically unfeasible despite many years of trials until K.G. Wilson [18] proposed to replace the continuum phase space by a discrete approximation, the size of the lattice introducing a cut-off stabilizing the solutions. It is only these recent years that this approach became successful with the tremendous increase in computational power of the super-computers and new breakthrough in the basic algorithms. This technique is known as the QCD calculus on lattice or lattice QCD. Please refer to the excellent status report by Karl Jansen in these proceedings.

## 3. From tree level to higher orders: the NLO Revolution

Perturbative field theory allows level by level calculation of particle interactions. The technique was developed by R.P. Feynman [19] in the 50's. Each precision level can be matched to a set of similar diagrams. Let's take a simple example, in QED[2] (fig. 4), at the lowest level or tree level in the coupling constant α, the simple $e^+e^- \to e^+e^-$ process involves only 2 diagrams: a s-channel graph when the electron and the positron annihilate to form a virtual photon, also called propagator, which then decay into an electron and a positron and a t-channel where a photon is exchanged between the electron and the positron. In QED, the coupling constant is small enough (~ 1/137, although also running) so that in most cases first order calculations provide good enough precisions matching experimental accuracies. The input and output particles are called "legs".

But in some cases, experimental errors are much smaller than tree level computations like at the precision $e^+e^-$colliders. They can even be orders of magnitude smaller when, for example measuring the muon anomalous magnetic moment factor $(g-2)/2$ at 0.14 ppm precision [20]. The theoretical precision should be, at least, as good and computing the next terms in the expansion series becomes necessary both for electroweak and hadronic corrections.

The next level in precision is the so-called "Next-to-Leading Order" (NLO). It essentially implies adding 1-loop corrections. A photon can be exchanged between the initial state electron and positron, an $e^+e^-$ loop can be inserted in the propagator or even (not shown), a second photon is exchanged in the t-channel (creating box diagrams). But the virtual exchanged particle can also be real and diagrams with an external photon must be included. If one sticks precisely to the same final state, the additional photon should not be observed namely, it should be either soft or almost collinear with the emitted electron. Loops and/or legs can be put on all elementary parts of the tree-level diagrams and the number of diagrams increases dramatically. Next-to-Next-Orders calculations follow the same trend, adding more and more diagrams and therefore becoming more and more computational intensive.

---

[1] QCD: Quantum ChromoDynamics, model of the strong interaction particularly probed at protons colliders like the LHC or at lepton-proton collider like HERA (DESY).

[2] QED: Quantum ElectroDynamics, quantum theory for the electromagnetic interactions.

At LHC, QCD plays a major role and things get more complicated. First the proton is not an elementary particle, but a complex system of 3 valence quarks in a gluon fog and a sea of quark-antiquark pairs. At LHC energy, the interest is on the hard scattering of proton constituents, namely gluon-gluon, quark-gluon or quark-quark. Therefore many different initial states must be taken into account. The selection of the initial partons within the proton is made through the use of the so-called parton distribution functions (PDF) which describe the parton content of the proton. These functions are mainly parameterizations based on data from the e-p experiments at HERA (DESY [21]) or in $p\bar{p}$ experiments at D0/CDF (Tevatron [22], Fermilab) and now at LHC.

Second, the final states may involve many particles, i.e. many legs, like shown in the experimental wish list discussed at the "Les Houches" meetings [23]. This is a huge step in complexity compared to the 2→4 channels studied at the electron-positron collider (LEP II [24]).

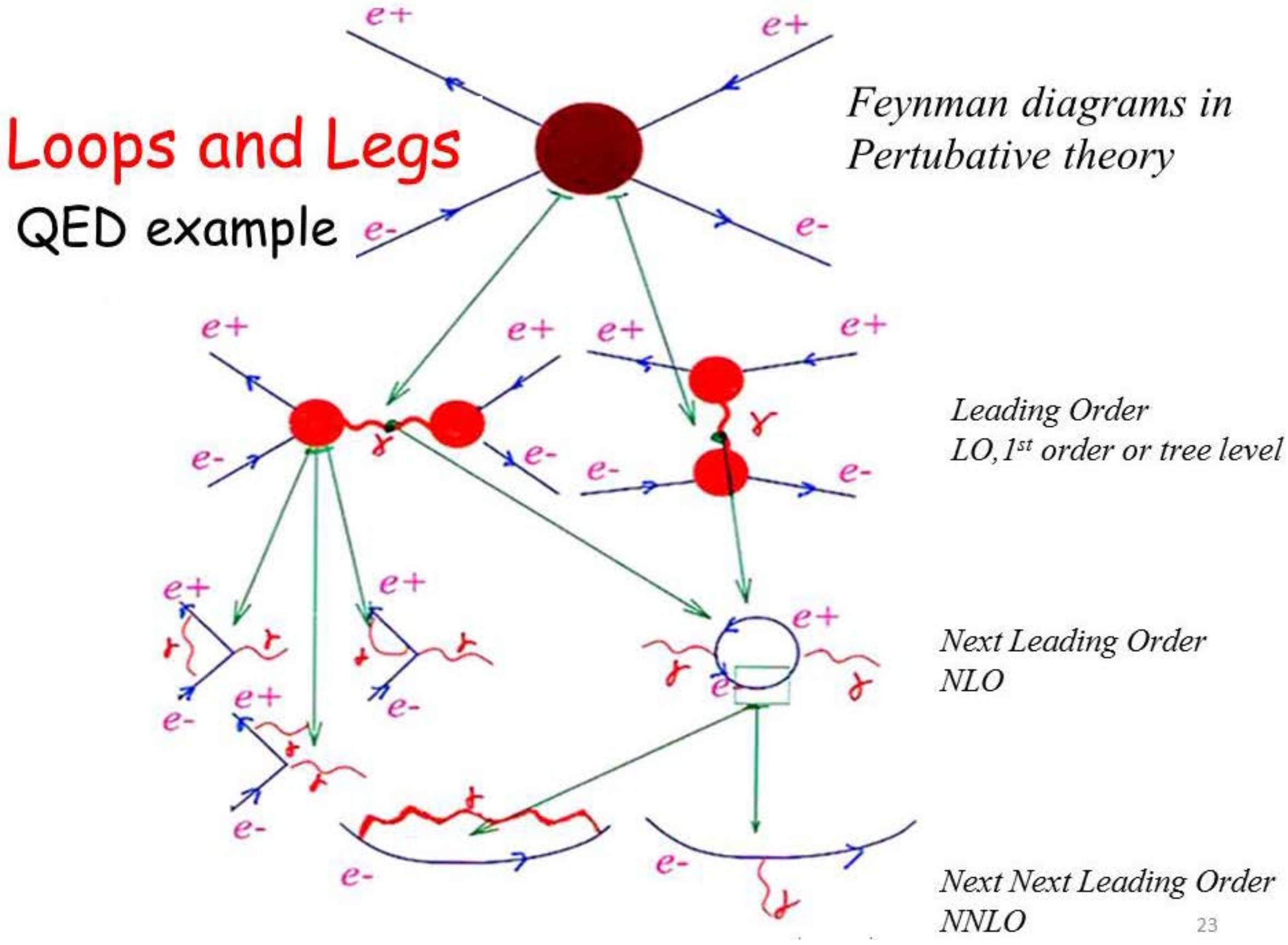

***Figure 4: Loops and legs in QED***

Finally, contrary to QED where tree-level calculations are often a good approximation, QCD estimations must go beyond LO which depend too much on arbitrary scales. The renormalization scale $\mu_R$ enters in the cross-section as well as in the value of $\alpha_S$. Renormalization is a necessary operation to cancel infinities arising in the calculation of the cross-section due to UV (Ultra-Violet) divergences. The factorization scale $\mu_F$ which also enters in the cross-section as well as in the parton distribution function is related to the soft and collinear divergences, namely in the IR (Infra-Red) region. These scales are arbitrary and induce large normalization errors (10-20% or even more). Higher order calculations reduce largely the sensitivity to these arbitrary scales and this trend is even more pronounced when the number of legs is large.

QCD NLO calculation of many legs processes may lead to more than $10^{5\text{-}6}$ diagrams and for each, the amplitude or matrix element can be so large in the usual Feynman representation that computer memory sizes become an issue. Clearly the Feynman diagrammatic approach was not quite suited for this kind of calculation.

Although this conventional method is still in use and even being improved, new approaches originated from the string/twistor theory have recently been developed for the calculation of one-loop processes. The on-shell unitarity method [25] to the expense of rewriting the quantum field theory in terms of invariant gauge on-shell intermediate expression simply proposes to cut the loop diagrams in one or many places decomposing the initial one-loop calculation to a set of tree level amplitudes, easily handled by the conventional tree-level diagram calculators. Following these ideas many developments have been undertaken (see for a summary until 2008 [26]) and have led to automatic calculation of 2→4 or 2→5 NLO processes (fig.5).

Even if these techniques for the most complex calculations have not reached full automation, they have made the computation possible with the current computers. Event generators at NLO based on SHERPA [27] and BlackHat [28], MadGraph [29] and others have been heavily used by the LHC experiments.

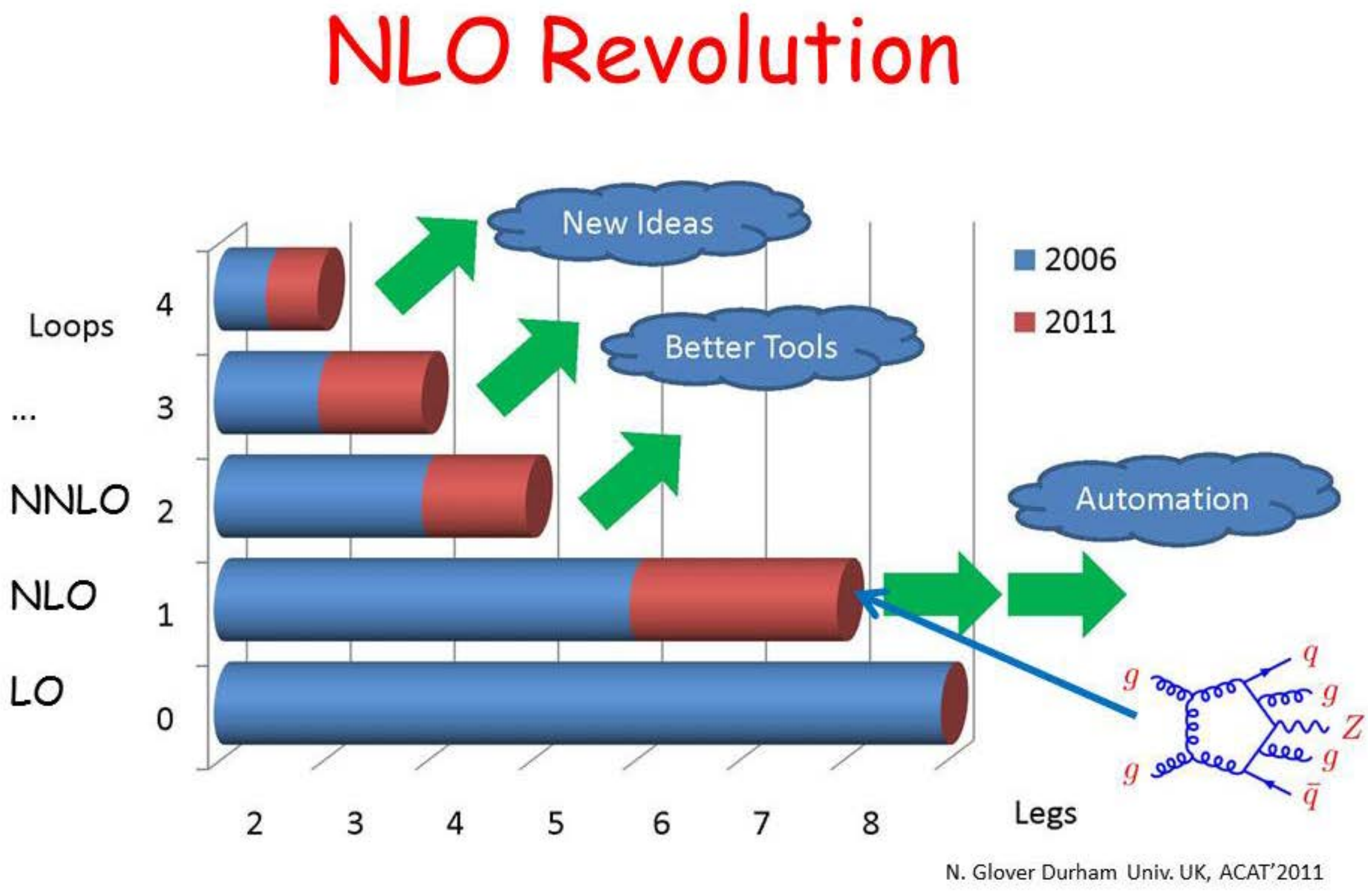

***Figure 5: NLO Revolution*** **[30]**

## 4. Event Simulation and data analysis

It is impossible to calculate analytically the detector responses to an event produced in the quite complex LHC experiments. One therefore relies on Monte-Carlo simulations. Any process of interest can be actually generated and simulated in a virtual representation of the detectors. The simulation is supposed to mimic in detail what is detected when beams collide.

Fig. 6 presents schematically the various steps of the simulation chain. On the top, the “physics model” describes the framework in which the calculation is performed. The basic model is the Standard Model (SM). But extension or beyond SM models are also implemented like for example the supersymmetric (SUSY) model. This model suggests that any known fermion (boson) has a partner boson (fermion) yet unobserved due to its high mass. Clearly beyond standard model (BSM) theories are made up to cure problems arising in the SM beyond current energy, in particular, the divergent contributions to the Higgs mass.

The actual process is, then, selected by defining the input and output particles as well as the order of calculation. The matrix element or the expression of the process probability is then automatically prepared based on the physics model data.

After integrating of the matrix element over the parameters phase space, the code for an event generator is constructed. It will generate random samples of energy-momentum four-vectors for all final state physical particles.

Each generated particle is then propagated into a model representing the experiment built by a detector simulation package. GEANT [31], initiated at CERN and developed by an international collaboration is the main package used by our community. Version 4 is an extremely detailed description of the detector components down to nuts and bolts and read-out cables. All physics particle-matter interactions have been implemented. Each particle, step by step is tracked in the material forming the structures as well as the detection sensitive elements. The charged particles may produce primary gas ionization, scintillation or Cherenkov light depending on the materials. Photons/gammas create electromagnetic showers producing hundred to thousand low energy particles which are all tracked down to final absorption. This information is collected by sensitive elements: wires, silicon strips/pixels or photo-detectors. All particles are followed until stopping in material or escaping the experimental set-up. The huge amount of information produced by the detector simulation should in principle represent exactly what a real event in a real detector would produce. Obviously, confidence on the detector simulation accuracy has been obtained after thousands of validation measurements for each type of particles, materials and detectors.

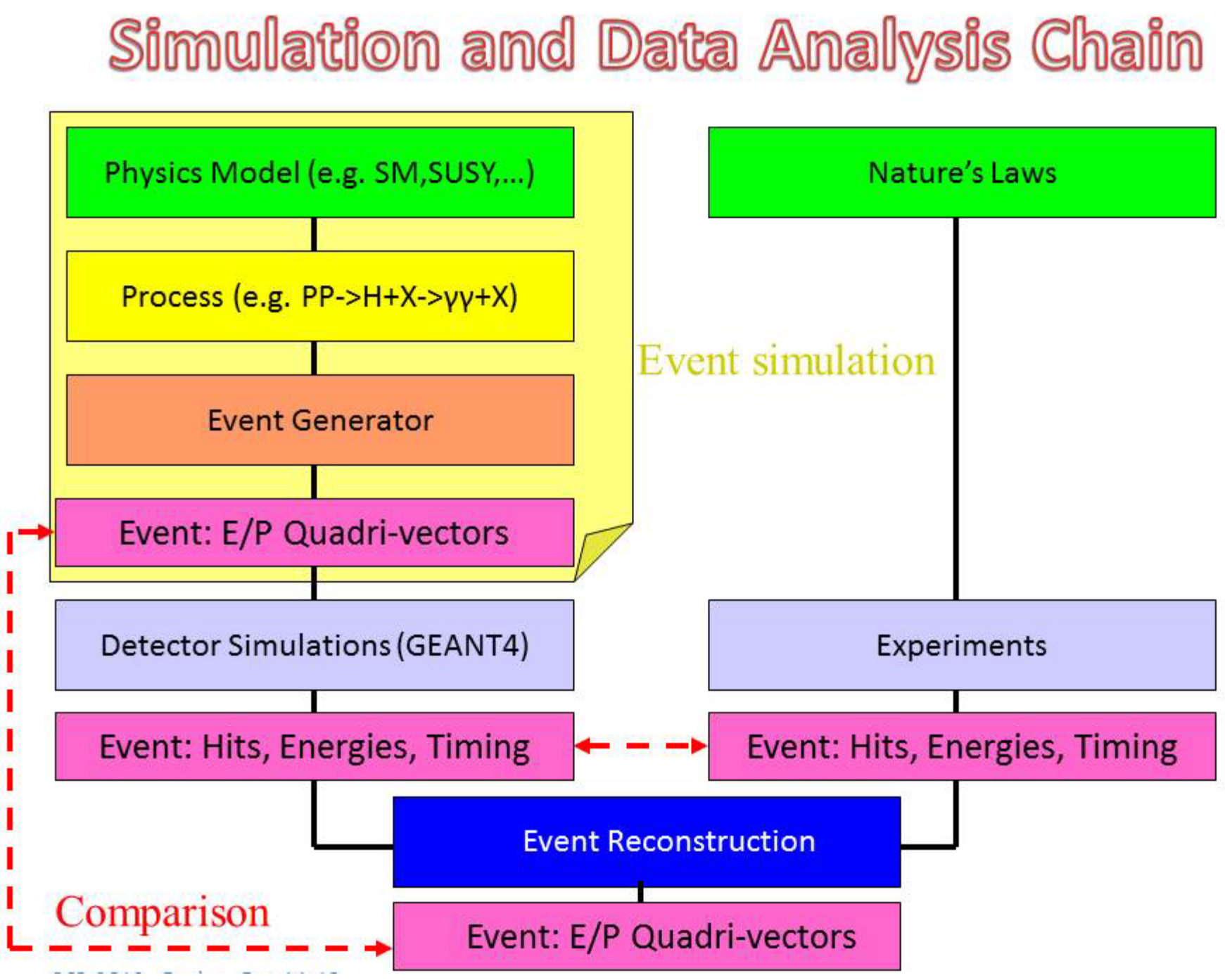

***Figure 6: Monte-Carlo simulation and the data analysis chain***

At this level, one has now simulated i.e. theory "generated" event data that can be compared to the experimental data produced by Nature in the physical detector. Both simulation and real data are then reconstructed by the same reconstruction programme to get the initial set of energy-momentum four-vectors representing the actual simulated or observed event. Comparing the reconstructed simulated four-momenta to those produced by the event generator is a test of self-consistency of the whole system although departure between the simulated and the experimental data may give hints to new physics discovery.

The final analysis is generally done with the Root data analysis toolkit [32], developed at CERN. Events data are put in large ntuples (sort of expandable spread sheets) on which filters and fits can be applied. Estimators can then be computed and placed in histograms. This is how plots like those shown in fig.3 have been produced. Would the experimental data have been precisely fitted by the green-yellow background band, we would had concluded to a no Higgs discovery and higher limits on its production would had been set. But hopefully some excess was found, signalling the existence of something beyond the expected background, a new particle, maybe the expected Higgs.

## 5. Event sequential formation stages

Let's now focus on the physics of an event formation. In a high-energy *pp* scattering, the generation of a single event (fig. 7) takes several steps. First, 2 partons (flavour and colour) are selected in each of the colliding protons. Before interacting, partons are allowed to emit gluons (initial state radiations). Then the actual hard-scattering interaction occurs and 2 or more particles are produced. These particles may then decay like in heavy quark/leptons or bosons (Z, W) producing several daughters. Finally, the last remaining quark and gluon undergo partonic showers before the final hadronization step occurs, namely the recombination of end partons in hadrons (baryons or mesons).

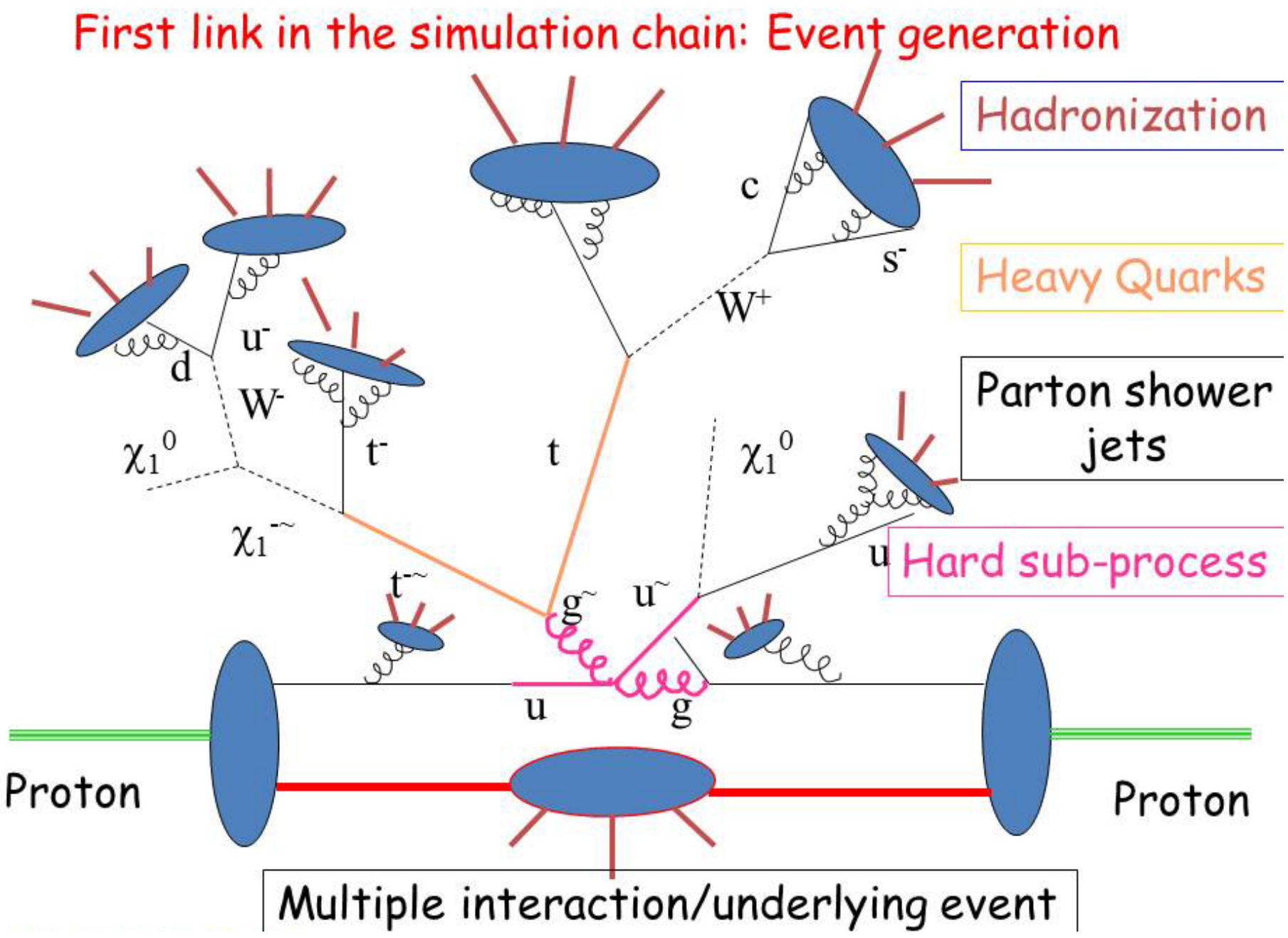

***Figure 7: Sequential steps in the formation of an event***

*(The event presented here is a SUSY event with a hard process: u-quark gluon →u-squark gluino, the u-squark being the superpartner of a u-quark and the gluino the gluon counterpart)*

In addition, several interactions between different partons can occur in the same proton collision leading to multiple parton interactions (MPI). Finally, in the collision of proton bunches more than one collision can occur and multiple vertices are created, this is called the underlying events (UE). The odds that several interactions producing a hard-scattering each leaving, in the detectors, high momentum transfer particles is quite small, but these extra collisions produce many additional tracks and energy depositions in the detector. These parasitic effects make the analysis more difficult as each track or energy deposition should be associated to the right vertex and, therefore, the whole reconstruction process takes more CPU resources and becomes more error prone.

## 6. Process calculation

Now that the physics of an event formation has been described, let us see how this is implemented in the global process calculation (fig. 8). A flurry of packages has been developed to carry out the full event simulation.

Based on the model Lagrangian or, more generally, on the group symmetries and fields definitions, symbolic manipulation languages like Reduce, Maple, Mathematica or Form [33] are used to encode in a model file the various couplings and the propagators expressions either on an analytic form or as a set of numerical libraries. The Form language has been developed to handle very large expressions buffered on disk files. Specific programs like LanHEP [34], FeynRules [35] or SARAH [36] (SUSY) have made this derivation automatic.

Then a specific process can be selected. The initial particles undergoing the hard collision are generated after initial state radiation and beamstrahlung (CIRCE [37]) for electrons or parton extraction from the parton distribution functions (PDF) (LHApdf [38], CTEQ [39], MRST [40], …) for the protons has been applied. One of the automatic amplitude/matrix element generators can be put in action to produce the global expression of the process integrand.

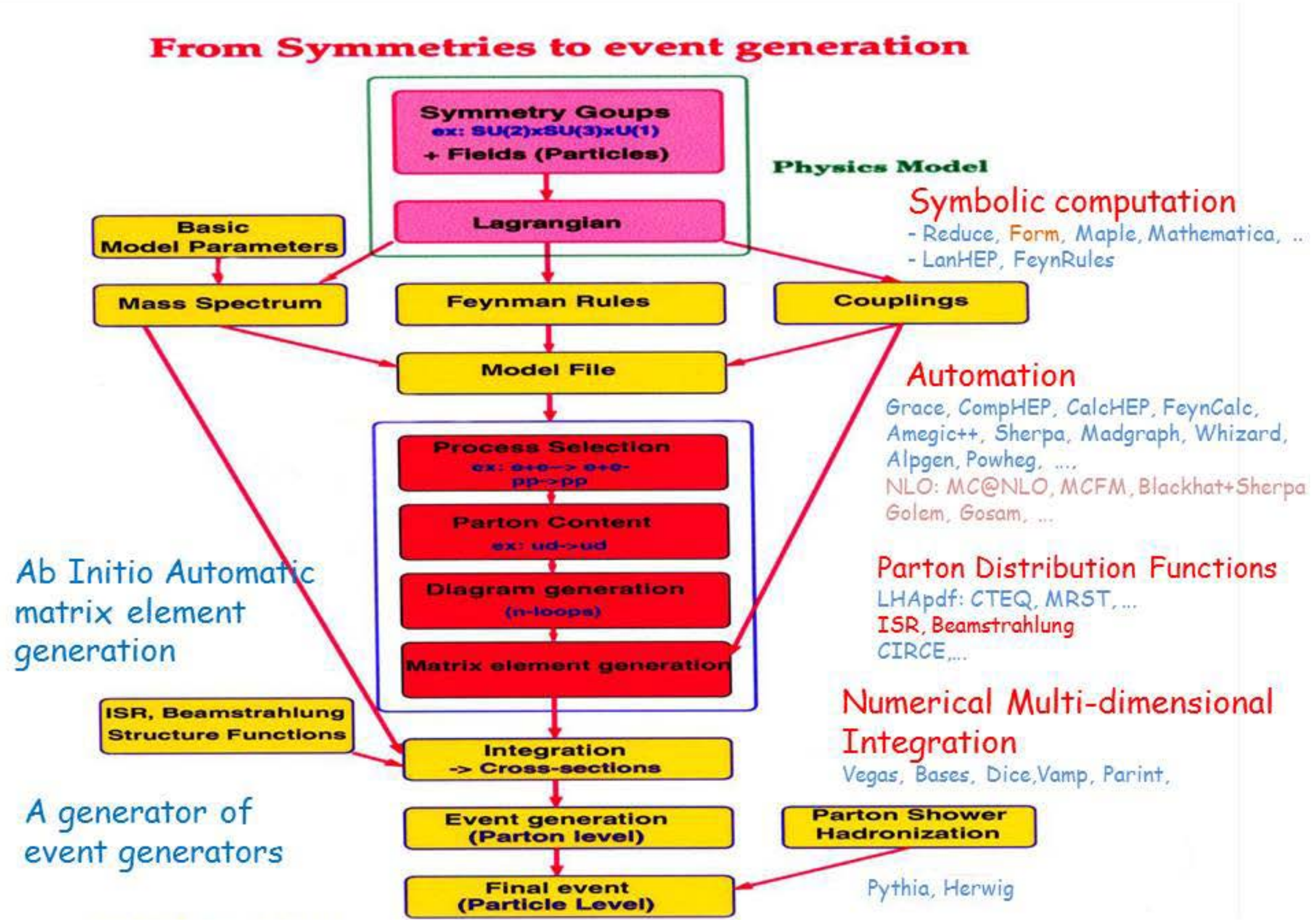

*Figure 8: From symmetries to Event Generation*

This expression must then be integrated over the full phase space, namely over all possible legs momenta/energy, only constrained by the energy momentum conservation and the initial conditions. If loops are involved, integration over the loop propagators must also be performed. The multi-dimensional integration provides the total cross-section and the probability function for the generation of the events.

The event generator can, now, be constructed. It will randomly generate events at the parton level. Parton shower and hadronization programs like Pythia [41] or Herwig [42] will take over to produce the final full-fledged event ready to enter the detector simulation stage.

## 7. Matrix element automatic calculations

The matrix element calculation it-self has been for many years performed manually by talented theorists. Back in the 80's, a group in Japan from the KEK laboratory (Grace) [43] and a Russian group from Moscow State University (CompHEP) [44] [45] started to develop packages to do these lengthy and error prone calculations automatically, that was the beginning of the automatic process calculation endeavour.

The automation was initially seen as a toy project, at best to train students. But the experimental requests for new process calculations went exploding as well as the complexity of the process calculation and, finally, the automatic approach became the baseline. The first event generator with an automated computed matrix element was probably grc4f $e^+e^-\rightarrow$ 4 fermions [46] which was used for the LEP2 analysis. For the first time, spin correlations and mass effects were included. Nowadays most calculations are made by computer tools in a combination of symbolic and numerical operations.

The various steps of the automatic calculation flowchart, at tree level can be summarized as follows:

1. Enumeration and description of the set of diagrams entering the calculation of a given process matrix element (e.g. standalone QGRAF [47] or included in the automatic packages e.g. [43], [44])
2. Preparation of the matrix element expression for each diagram by various calls to the vertex and propagator subroutine library or by building a global analytical expression.
3. Multi-dimensional integration over the full phase space (including the PDF variables) to produce 1) the total cross-section 2) the probability function describing how likely an event with a given set of input and output four-vector will occur.
4. Preparation of the event generator code. According to the probability function, random set of the process four vectors will be generated.
5. Inclusion of the final state radiation calls as well as the parton showering and hadronization steps based on Pythia [41] or Herwig [42].

For reference here follows some tree level matrix element generators: GRACE [43], CompHEP [44], CalcHEP [48], FeynCalc [49], MadGraph [29], AMEGIC++ [50] as well as multi-channel event generators HELAC-PHEGAS [51], ALPGEN [52], GR@PPA [53], O'Mega [54], Sherpa [27], Whizard [55], Pythia [41], Herwig [42].

In practice there are, yet, no complete and fully automatic matrix element generators at NLO, but many packages are being developed in this direction. Recent reports can be found in Ref. [56,57,58] and the main contenders are: MCFM [59], NLOJET++ [60], BlackHat [61], Rocket [62], CutTools [63], MadLoop [64], OpenLoops [65], GOLEM [66], POWHEG [67], aMC@NLO [68,69], Sherpa+BlackHat [70], MadGOLEM [71], GoSam [72], HELAC-NLO [73].

## 8. Matrix element multi-dimensional Monte-Carlo integration

Unfortunately matrix elements tend to have singularities in some parts of the phase space making their integration difficult: for example, in the collinear region where two particles (e.g. an electron and a photon) get a small angular separation. This corresponds to the quantum physical situation where an electron alone cannot be distinguished from an electron closely accompanied by a soft photon. In that case one gets a singularity that can be tamed by introducing a cut on the photon-electron angle smaller than the actual experimental angular resolution.

Anyway singularities occur in the expression of the integrand and the skill of the event generator developers are still needed to craft the most efficient variable transformations so that singularities are avoided or minimized. The mapping between integration and physical variables is coded into the "kinematics subroutines". They are selectively plugged into the integrand code according to its sensitivity to specific divergences.

For the less acute variations, different techniques automatically adapting the sampling grid to the shape of the integrand have been implemented in the main multi-dimensional integration packages like Vegas [74,75], Bases/Spring [76], Dice [77], Foam [78], Vamp [79], Mint [80], ParInt [81]:

- Importance sampling: the grid intervals or bins get smaller and therefore the sampling is denser in regions of high integrand values as this phase space regions will contribute most to the total integral.
- Stratified sampling: the bins get smaller in the region of fast variation of the integrand so that from one interval to the next the function varies slowly and smoothly.
- Multi-channel sampling: several mappings, namely grid definitions, can be selected depending on the phase space region. The method is to always select the mapping that suits best the behavior of the integrand at the particular sampling point.

For processes up to $2 \rightarrow 4$, at tree level the full automation is now usual business. For more particles in the final state (e.g. $2 \rightarrow 6$), the integration stage may take a lot of CPU time and the calculation may become intractable if the kinematics mappings are not adequate.

For NLO calculations the number of integration dimensions increases due to the additional loop four-momenta and the integration difficulties become even more stringent. In addition, the size of the integrant can be enormous and compiler and computer memory limitations become an issue. Fortunately after the NLO revolution, some computations have become feasible, but automation has not yet been fully implemented.

When the integrand becomes too large, it can be cut in smaller parts (e.g. each sub-diagram) distributed over many nodes. This is one direction for parallelization. A second one to accelerate the computation is to share the whole phase space again over many nodes. So that, in the end, each node will compute a part of the integrand for a part of the phase space. However, the integration algorithm requires the full value of one or several integrand calculations over the whole phase space, before executing the next iteration. This is the serial part of the integration algorithm and load balancing between the nodes becomes an issue that has to be addressed [82,83] .

Recently, it has been shown that the computation of loop integrals or tree level processes on GPUs have led to dramatic acceleration [84,85]. Supercomputers will certainly benefit these calculations.

## 9. Matrix element method for data analysis.

As we have seen, the process matrix element is at the heart of the event generator and its calculation can be made automatic in most cases, at least, at tree level. But a new application of the matrix element is looming up for the data analysis it-self.

The usual way of extracting new insights from the experiments is by comparing the simulated and experimental event distributions. This is the conventional approach. A more elaborate, but similar technique called the template analysis has been developed along these lines and involving likelihood maximization. But a new approach is nowadays preferred. It involves the use of the matrix element expression in the event selection algorithm it-self. This is the matrix-element method [86] or also called the matrix element likelihood algorithm (MELA) in CMS.

### 9.1. Template method

Let us assume that a process parameter is to be estimated from experimental data. Let's take for example, the mass of the top quark. The t-quark has a very short life-time and decays, even before it gets hadronized, in various channels including in lepton + jet. From these decay products the top quark mass can be reconstructed. The template method requires the simulation of several similar distributions assuming different top masses. Then, a likelihood expression is built and its maximization provides the best fit between the experimental and simulated distributions. This method is straight forward, but only a small part of the full event information is used, namely the reconstructed top mass. The rest of the event data like the distributions of remaining particles and their correlations does not enter this optimization.

9.2. Matrix-element method

From the same selected sample, a probability is built for each event taking into account all final state particle kinematics to estimate to which category it is most likely part of. The following probability $P_{sig}$ is therefore estimated event per event:

$$P_{sig}(x, m_t) = \frac{1}{N}\int_{y,q_1,q_2} d\Phi(y)\ |M_{t\bar{t}}(y, m_t)|^2\ W(x, y)\ f_{pdf}(q_1)\ f_{pdf}(q_2)\ d\,q_1 dq_2$$

Where:

$x$ represents the measured kinematic variables for all final state particles

$m_t$ is the parameter to be estimated, here the top mass

$|M_{t\bar{t}}|^2$ is the square of the matrix element associated to the signal process under study, here the production of a pair of top quarks.

$W(x, y)$ is the transfer function mapping the measured four-vectors and the matrix element theoretical parameters.

$f_{pdf}(q)$ is the parton distribution function described above giving the probability that the initial colliding partons with $q_1$ and $q_2$ four-momenta are produced by the interacting protons.

This expression is integrated over the phase space and normalized to the observed cross-section $N$.

Similar probabilities $P_{bkg1\dots n}$ are formed assuming that the same event belongs to background processes, namely processes having a similar final state but not through the decay of top quarks.

If the $f_{sig,bkg1\dots n}$ are the fractional parts of the total sample (signal or backgrounds), a global probability for one event with the kinematics $x$ to form a mass $m_t$ whether from signal or from background can then be computed:

$$P_{evt}(x, m_t) = f_{sig}\ P_{sig} + \ f_{bkg1}\ P_{bkg1} + \cdots + \ f_{bkgn}\ P_{bkgn}$$

Finally, a sample level likelihood is computed as the product of the global probability of all events and the following inverse log is minimized to get the most probable $m_t$:

$$\mathcal{L}(m_t) = \ -ln(\textstyle\prod_{evt} P_{evt}(x, m_t))$$

In this method implemented in the MadWeight [87] package, the whole event information is incorporated in the parameter estimation. In addition the separation between the experimental data and the theory input is well identified.

But this is a very CPU time consuming procedure as for each event several multi-dimensional integrations are necessary. To help solving this problem, the use of GPU has been implemented successfully [88].

The template (fig. 9) and the matrix element method (fig.10) have been used with great success for establishing the value of the top mass at the Tevatron by the D0 [89] and CDF [89] experiments.

A similar technique has also been used at LHC to estimate the Higgs mass. A global expression MELA is computed from the likelihood, namely MELA is 0 (1) if the probability of the event to be a background (signal) is maximal. Each event is presented on a scatter plot with the estimated Higgs mass on the x axis and the MELA value on y-axis. Setting the threshold on MELA above 0.5 selects events more likely to be a signal than a background which greatly enhances the signal over noise ratio of the Higgs mass measurement (fig. 11).

Matrix element methods incorporating NLO corrections have also been studied [90], but have not yet been implemented for experimental data analyses.

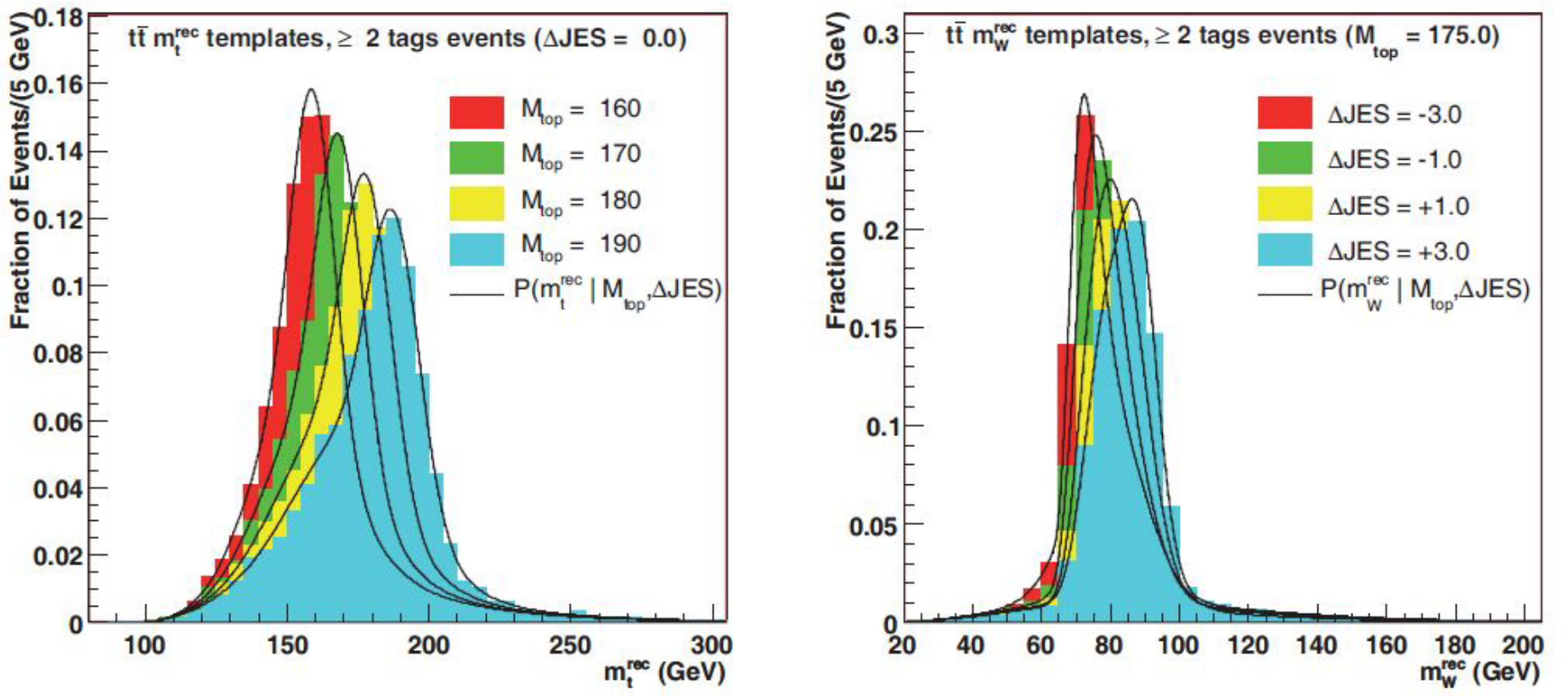

*Figure 9: The template distributions for reconstructed top mass (left) and mw (right) for CDF* [89]

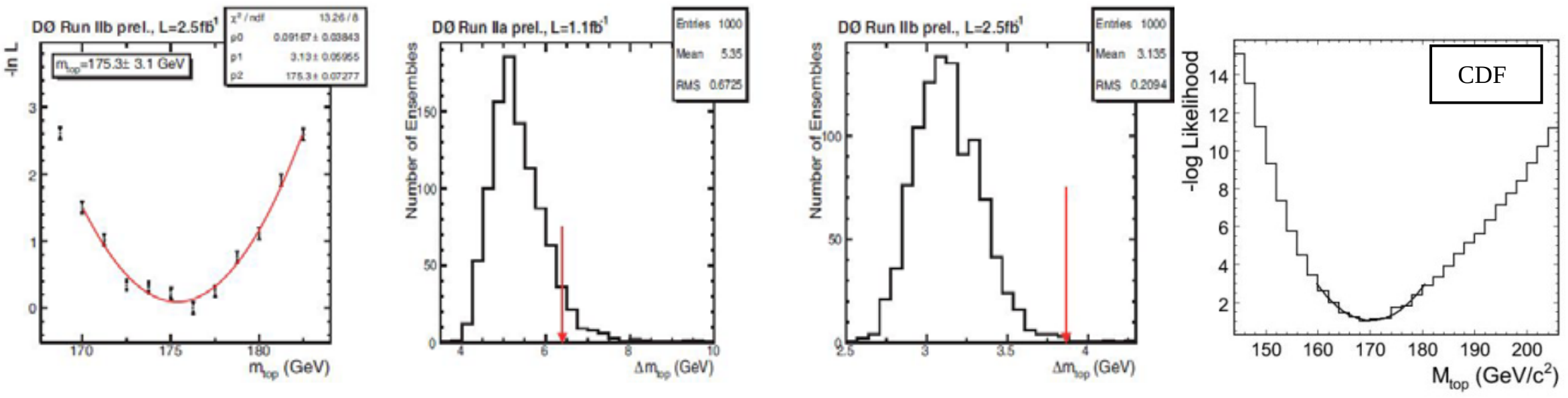

***Figure 10: Measured likelihood for the top mass at D0 and CDF* [89,91]**

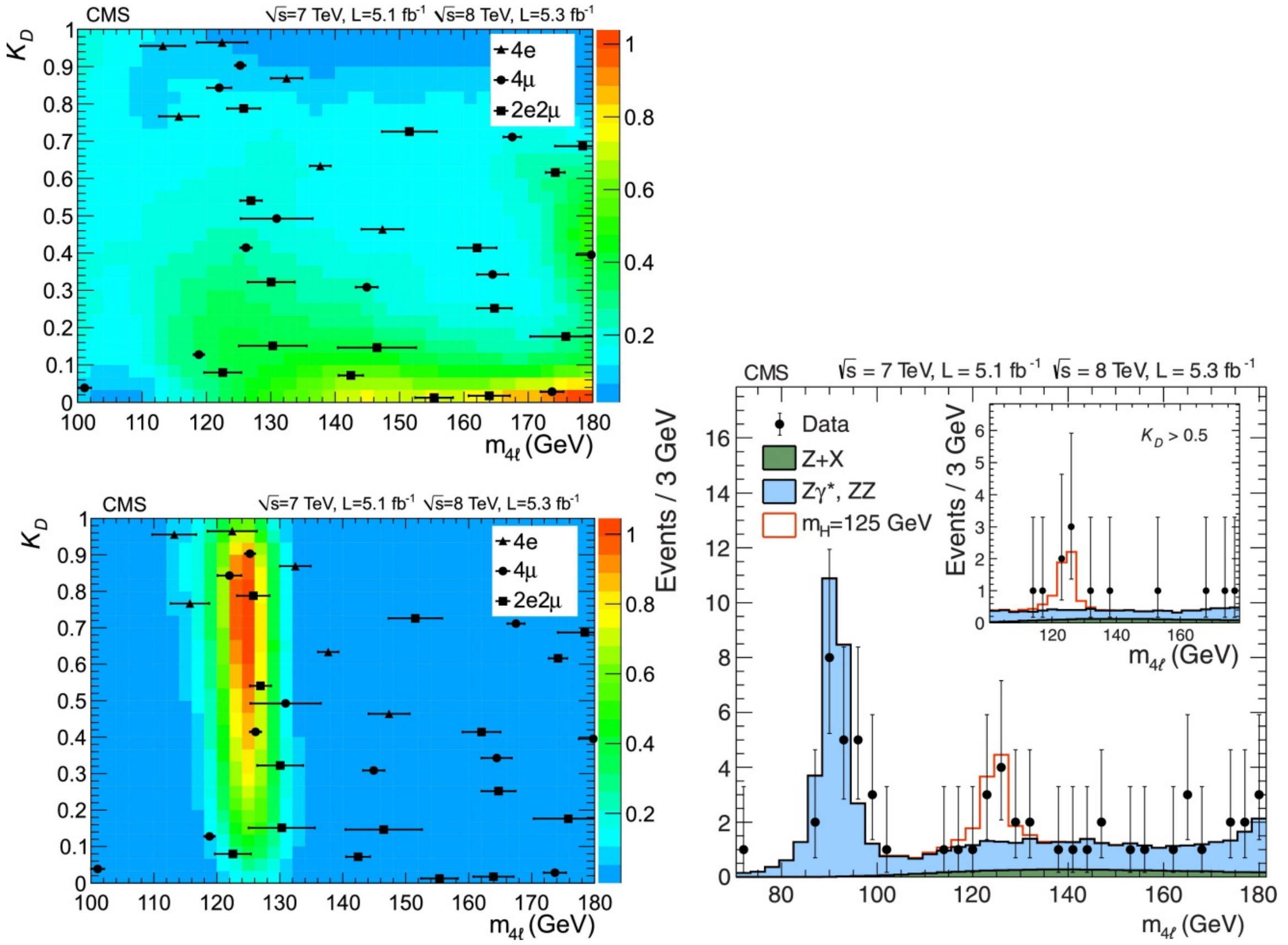

***Figure 11: CMS: MELA analysis of Higgs decaying in 4 leptons (from* [11]*)***

## 10. Cross-disciplinary fertilization and conclusions

Computational particle physics covers two main branches, one dealing with the non-perturbative calculations, lattice QCD, which is virtually intractable without the use of large parallel super computers. Recent developments show that thanks to a new algorithm and a drastic increase in computing power, the hadronic mass spectrum and other global parameters are now well reproduced.

The other branch is related to the perturbative region probing the inner part of the nucleon as well as its high energy dynamics. These computations, for the highest available energy, have also become feasible thanks to new NLO calculation methods and, again, to the availability of a large Grid computing infrastructure. The automation is now on track for the most complex calculations. Experimental design, physics analysis and theoretical interpretation at current colliders deeply rely on the development of these "new" tools providing matrix elements and event generators.

In fact, perturbative calculations both benefit from and motivate the development of techniques and methods common to other domains in computational science including:

- Symbolic calculation: the traduction of symmetries and fields to Feynman rules and couplings has been automated using algebraic languages like Reduce, Maple or Mathematica. Often created to support these calculations, they are now multi-purpose languages and are heavily used in many other research fields. However, the handling of very large expression of high complexity has triggered the implementation of expression disk caching and parallel techniques like in the language Form.
- Parallelism:
    - Numerical multi-dimensional integration: one of the main issues in perturbative computation is the integration over many dimensions of very large singular expressions. Some of the integration packages have been made parallel and able to run on GPU or supercomputers.
    - Process calculation factorial growth: the number of models, of processes and sub-processes is sky rocketing prompting for the use of massively parallel supercomputers or large clusters.
- High precision floating point computation: often double precision is not sufficient. Quadruple or octuple precision becomes necessary to achieve convergence for the most singular integrands leading to order of magnitude increase in computing time [92]. Specific libraries and hardware developments based on co-processors or GPUs [93] are in progress.
- Large computing infrastructures:
    - The internationally distributed Data Grid architecture, mostly created to cover the collider experiments needs, has fulfilled the expectations. It gave birth to the "Cloud" technology, a new commercial endeavor.
    - Although most of the current data analysis and simulations programs are serial, both the many-core/GPU long term industrial trends and the need for higher precision theoretical predictions are incentives to implement multi-threading and parallel execution techniques (e.g. for analysis [94] and simulation [95]). Certainly not an easy task in object oriented environments, but it will open access to the huge power of supercomputers.
- General data analysis toolkits (e.g. Root) implementing advanced statistical methods developed for particle physics has found applications in other environment like biology, finance and medicine.

Finally, the initial dream of automatic perturbative physics calculations has triggered the development of a flurry of packages and specific tools. Some are at the heart of today high-energy colliders and astroparticle experiment data analysis. But these packages have usually very little in common. Despite great attempts to standardize some aspects like the output files format (StdHep [96], HepMC [97]) or the interface with NLO calculations (BLHA [98]), they still have no common input syntax, no general parameter definition format, a poor modularity, no universal submission scheme, no common visualization and analysis tools, and so on. It is time to address these issues to facilitate their necessary detailed comparisons and their embedding in the large experimental simulation packages.

**References:**

[1] D. Perret-Gallix. CCP 2012: Computational particle physics for event generators and data analysis (slides). [Online]. http://acpp-coll.in2p3.fr/cgi-bin/twiki.source/pub/ACPP/PresentationsNotes/ccp_2012-3.pptx

[2] LHC. Large Hadron Collider. [Online]. http://public.web.cern.ch/public/en/LHC/LHC-en.html

[3] CERN. The European Center for Nuclear Research. [Online]. http://www.cern.ch

[4] Pierre Auger Observatory. [Online]. http://www.auger.org/

[5] GEM-EUSO. proposal for Gem-Euso. [Online]. http://jemeuso.riken.jp/en/index.html

[6] ATLAS experiment at CERN LHC. [Online]. http://atlas.ch/

[7] CMS experiment at CERN LHC. [Online]. http://cms.web.cern.ch/

[8] LHC-B experiment at CERN LHC. [Online]. http://lhcb.web.cern.ch/lhcb/

[9] ALICE experiment at CERN LHC. [Online]. http://aliceinfo.cern.ch/

[10] ATLAS Collaboration, "Observation of a new particle in the search for the Standard Model Higgs boson with the ATLAS detector at the LHC," *Phys. Lett. B*, vol. 716, pp. 1-29, 2012. [Online]. http://arxiv.org/abs/1207.7214

[11] CMS Collaboration, "Observation of a new boson at a mass of 125 GeV with the CMS experiment at the LHC," *Physics Letters B*, vol. 716, no. 1, pp. 30-61, 2012. [Online]. http://arxiv.org/abs/1207.7235

[12] Missing Higgs. [Online]. http://public.web.cern.ch/public/en/science/higgs-en.html

[13] Brout Robert and Englert Francois, "Broken Symmetry and the Mass of Gauge Vector Mesons," *Physical Review Letters*, vol. 13, no. 9, pp. 321-323, 1964.

[14] Peter Higgs, "Broken Symmetries and the Masses of Gauge Bosons," *Physical Review Letters*, vol. 13, no. 16, pp. 508–509, 1964. [Online]. http://prl.aps.org/pdf/PRL/v13/i16/p508_1

[15] Gerald Guralnik, Hagen, C. R. Hagen, and T. W. B. Kibble, "Global Conservation Laws and Massless Particles," *Physical Review Letters*, vol. 13, no. 20, pp. 585–587, 1964.

[16] WLCG: Worldwide LHC Computing GRID. [Online]. http://wlcg.web.cern.ch/

[17] David Kaiser, "Physics and Feynman's Diagrams," *American Scientist*, vol. 93, p. 156, 2005. [Online]. http://web.mit.edu/dikaiser/www/FdsAmSci.pdf

[18] Kenneth G. Wilson, "Confinement of quarks," *Physical Review D*, vol. 10, no. 8, p. 2445.

[19] R.P. Feynman, "Space-time approach to quantum electrodunamics," *Physics Review*, vol. 76, no. 6, pp. 769-789, 1949.

[20] Fermilab E-898, "The New (g – 2) Experiment: A proposal to measure the Muon Anomalous Magnetic moment to +- 14 ppm precision". [Online]. http://gm2.fnal.gov/public_docs/proposals/Proposal-APR5-Final.pdf

[21] Hera at Desy (Hamburg). [Online]. http://www.desy.de/research/facilities/hera_experiments/index_eng.html

[22] Tevatron at fermilab \. [Online]. http://www.fnal.gov/pub/science/experiments/energy/tevatron/

[23] "Les Houches" Working group, "THE SM AND NLO MULTILEG AND SM MC WORKING GROUPS," 2010. [Online]. http://arxiv.org/pdf/1203.6803.pdf

[24] LEP at CERN (Geneva). [Online]. http://public.web.cern.ch/public/en/research/lep-en.html

[25] L. J. Dixon, D. C. Dunbar, D. A. Kosower Z. Bern, "Fusing gauge theory tree amplitudes into loop amplitudes," *Nucl. Phys. B*, vol. 435, pp. 59–101, 1995.

[26] "Les Houches" Working group, "THE NLO MULTILEG WORKING GROUP: Summary Report," 2008. [Online]. http://arxiv.org/pdf/0803.0494.pdf

[27] SHERPA collaboration, "SHERPA". [Online]. http://sherpa.hepforge.org/trac/wiki

[28] Z. Bern, G. Diana, L. Dixon, F. Febres Cordero, S. Hoeche et al., "Four-Jet Production at the Large Hadron Collider at Next-to-Leading Order in QCD," 2011. [Online]. http://arxiv.org/pdf/1112.3940.pdf

[29] Michel Herquet, Fabio Maltoni, Olivier Mattelaer, Tim Stelzer Johan Alwall, "MadGraph 5 : Going Beyond," *JHEP*, vol. 1106, p. 128, 2011. [Online]. http://madgraph.hep.uiuc.edu/

[30] Nigel Glover, Gudrun Heinrich, "Computations in Particle Physics:," in *ACAT' 2011 Uxbridge, London*

*UK (Advanced Computing and Analysis technologies)*, 2011. [Online]. http://indico.cern.ch/getFile.py/access?contribId=124&sessionId=15&resId=0&materialId=slides&confId=93877

[31] Geant 4 collaboration. Geant 4. [Online]. http://geant4.cern.ch/

[32] Root Collaboration. Root. [Online]. http://root.cern.ch/drupal/

[33] T. Ueda, J.A.M. Vermaseren, J. Vollinga. J. Kuipers, "FORM version 4.0.". [Online]. http://arxiv.org/abs/arXiv:1203.6543

[34] A, Semenov, "LanHEP - a package for automatic generation of Feynman rules from the Lagrangian. Updated version 3.1". [Online]. http://arxiv.org/abs/arXiv:1005.1909

[35] C. Duhr N. D. Christensen, "FeynRules - s - Feynman rules made easy," *Comput. Phys. Commun.*, vol. 180, pp. 1614-1641, 2009. [Online]. http://arxiv.org/abs/arXiv:0806.4194

[36] F. Staub, "Sarah". [Online]. http://arxiv.org/abs/arXiv:0806.0538

[37] T. Ohl. Beam spectra: CIRCE, Luminous, Pandora. [Online]. http://theorie.physik.uni-wuerzburg.de/~ohl/lc/generators-sections013.html#circe

[38] Andy Buckley Mike Whalley. LHAPDF. [Online]. http://lhapdf.hepforge.org/

[39] CTEQ collaboration. The Coordinated Theoretical-Experimental Project on QCD. [Online]. http://www.phys.psu.edu/~cteq/

[40] James Stirling, Robert Thorne, Graeme Watt Alan Martin. MRS/MRST/MSTW Parton distributions. [Online]. http://durpdg.dur.ac.uk/hepdata/mrs.html

[41] Torbjörn Sjöstrand et al. Pythia. [Online]. http://home.thep.lu.se/~torbjorn/Pythia.html

[42] B. Webber et al. HERWIG. [Online]. http://www.hep.phy.cam.ac.uk/theory/webber/Herwig/

[43] T. Kaneko, S. Kawabata, and Y. Shimizu, "Automatic generation of Feynman graphs and amplitudes in QED," *Computer Physics Communications*, vol. 43, no. 2, pp. 279-295, January 1997. [Online]. http://minami-home.kek.jp/

[44] E. E. Boos et al., "ComHEP: Computer system for calculations of particle collision characteristics at High Energy," Moscou State University: Institute of Nuclear Physics, Preprint 89-63/140 [web site] http://comphep.sinp.msu.ru/, 1989. [Online]. http://ccdb5fs.kek.jp/cgi-bin/img/allpdf?199003342

[45] M N Dubinin E E Boos, "Problems of automatic calculation for collider physics," *Physics Uspekhi*, vol. 53, no. 10, pp. 1039-1051, 2010.

[46] T. Ishikawa, T. Kaneko, K. Kato, S. Kawabata, Y. Kurihara, T. Munehisa, D. Perret-Gallix, Y. Shimizu, H. Tanaka J. Fujimoto, "Grc4f v1.1: A Four fermion event generator for e+ e- collisions," *Comput.Phys.Commun.*, vol. 100, pp. 128-156, 1997.

[47] P. Nogueira, "Automatic Feynman graph generation," *Journal of Computational Physics*, vol. 105, pp. 279-289., 1993. [Online]. http://cfif.ist.utl.pt/~paulo/qgraf.html

[48] A. Puhkov. CalcHEP. [Online]. http://theory.sinp.msu.ru/~pukhov/calchep.html

[49] R. Mertig. FeynCalc. [Online]. http://www.feyncalc.org/

[50] R. Kuhn, G. Soff F. Krauss, "AMEGIC++ 1.0, A Matrix Element Generator In C++," *JHEP 0202:044,2002*, vol. 0202, p. 044, 2002.

[51] Aggeliki Kanaki, Costas G. Papadopoulos, "HELAC-PHEGAS: automatic computation of helicity amplitudes and cross sections". [Online]. http://arxiv.org/abs/hep-ph/0012004

[52] Mauro Moretti, Fulvio Piccinini, Roberto Pittau and Antonello Polosa Michelangelo L. Mangano. ALPGEN V2.14: A collection of codes for the generation of multi-parton processes in hadronic collisions. [Online]. http://mlm.home.cern.ch/mlm/alpgen/

[53] S. Odaka, Y. Kurihara, "GR@PPA 2.8: initial-state jet matching for weak boson production processes at hadron collisions," *Comput. Phys. Commun.*, vol. 183, p. 1014. [Online]. http://arxiv.org/abs/1107.4467

[54] T. Ohl et al. O'Mega: An Optimizing Matrix Element Generator. [Online]. http://theorie.physik.uni-wuerzburg.de/~ohl/omega/

[55] Thorsten Ohl, Jürgen Reuter, and Christian Speckner. Wolfgang Kilian. The WHIZARD Event Generator. [Online]. http://whizard.hepforge.org/

[56] Thomas Binoth, "LHC phenomenology at next-to-leading order QCD: theoretical progress and new

results," , 2008, p. POS (ACAT08) 011. [Online]. http://pos.sissa.it/archive/conferences/070/011/ACAT08_011.pdf

[57] Daniel Maitre, "Automation of Multi-leg One-loop virtual Amplitudes," , 2010, p. POS (ACAT2010)008. [Online]. http://pos.sissa.it/archive/conferences/093/008/ACAT2010_008.pdf

[58] Francesco Tramontano, "Progress in automated Next-to-Leading-Order calculations," in *ACAT 2011: Advanced Computing and Analysis Technologies for Research*, vol. 368 (2012) 012058. [Online]. http://iopscience.iop.org/1742-6596/368/1/012058/pdf/1742-6596_368_1_012058.pdf

[59] Keith Ellis, Ciaran Williams John Campbell. MCFM - Monte Carlo for FeMtobarn processes. [Online]. http://mcfm.fnal.gov/

[60] Zoltan Nagy, "Next-to-leading order calculation of three jet observables in hadron hadron collision.," *Phys.Rev. D*, vol. 68, p. 094002, 2003. [Online]. http://www.desy.de/~znagy/Site/NLOJet++.html

[61] C.F. Berger, Z. Bern, Lance J. Dixon, F. Febres Cordero, D. Forde, H. Ita, D.A. Kosower, D. Maitre, "One-Loop Calculations with BlackHat," *Nucl.Phys.Proc.Suppl.*, vol. 183, pp. 313-319, 2008. [Online]. http://arxiv.org/abs/arXiv:0807.3705

[62] Ellis R K, Giele W T, Kunszt Z, Melnikov K and Zanderighi G, "One-loop amplitudes for W + 3 jet production in hadron collisions," *JHEP*, vol. 0901, p. 012, 2009. [Online]. http://arxiv.org/pdf/0810.2762.pdf

[63] G. Ossola, C. G. Papadopoulos and R. Pittau, *JHEP*, vol. 0803, p. 042, 2008. [Online]. http://www.ugr.es/~pittau/CutTools/

[64] V. Hirschi, R. Frederix, S. Frixione, M. V. Garzelli, F. Maltoni and R. Pittau, "Automation of one-loop QCD corrections," *JHEP*, vol. 1105, p. 044, 2011. [Online]. http://arxiv.org/abs/1103.0621

[65] F. Cascioli, P. Maierhoefer, S. Pozzorini, "Scattering Amplitudes with Open Loops," *Phys. Rev. Lett.*, vol. 108, p. 111601, 2012.

[66] G. Cullen et al., "Recent Progress in the Golem Project". [Online]. http://arxiv.org/pdf/1007.3580.pdf

[67] Carlo Oleari. the PowHeg box. [Online]. http://arxiv.org/abs/1007.3893

[68] S. Frixione, B.R. Webber, "Matching NLO QCD computations and parton shower simulations". [Online]. http://arxiv.org/abs/hep-ph/0204244

[69] J. Alwall et al. aMC@NLO. [Online]. http://amcatnlo.web.cern.ch/amcatnlo/

[70] Kemal Ozeren, Lance J. Dixon, Stefan Hoeche, Fernando Febres Cordero, Harald Ita, David Kosower, Daniel Maître Zvi Bern, "High multiplicity processes at NLO with BlackHat and Sherpa," *Proceedings of Loops and Legs 2012*. [Online]. http://arxiv.org/pdf/1210.6684.pdf

[71] David Lopez-Val, Dorival Goncalves-Netto, Kentarou Mawatari, Tilman Plehn, Ioan Wigmore. MadGolem: automating NLO calculations for New Physics. [Online]. http://arxiv.org/abs/1209.2797

[72] Nicolas Greiner, Gudrun Heinrich, Gionata Luisoni, Pierpaolo Mastrolia, Giovanni Ossola, Thomas Reiter, Francesco Tramontano Gavin Cullen. GoSam. [Online]. http://gosam.hepforge.org/

[73] G. Bevilacqua, M. Czakon, M.V. Garzelli, A. van Hameren, A. Kardos, C.G. Papadopoulos, R. Pittau, M. Worek, "HELAC-NLO". [Online]. http://arxiv.org/abs/1110.1499

[74] G.P. Lepage, "A New Algorithm for Adaptive Multidimensional Integration, ," *Journal of Computational Physics*, vol. 27, pp. 192–203, 1978.

[75] VEGAS - GNU Scientific Library -- Reference Manual. [Online]. http://www.gnu.org/software/gsl/manual/html_node/VEGAS.html

[76] S. Kawabata, "A new version of the multi-dimensional integration and event ," *Computer Physics Communications*, vol. 88, pp. 309-326, 1995.

[77] K. Tobimatsu, S. Kawabata F. Yuasa, "Parallelization of the multidimensional integration," *Nuclear Instruments and Methods in Physics Research A*, vol. 502, pp. 599–601, 2003.

[78] S. Jadach, "Foam: Multi-Dimensional General Purpose Monte Carlo Generator With Self-Adapting Simplical Grid," *Comput. Phys. Commun.*, vol. 130, pp. 244-259, 2000. [Online]. http://arxiv.org/abs/physics/9910004

[79] T. Ohl, "VAMP, Version 1.0: Vegas AMPlified: Anisotropy, Multi-channel sampling and Parallelization," 1999. [Online]. http://whizard.hepforge.org/vamp.pdf

[80] P. Nason, "MINT: a Computer Program for Adaptive Monte Carlo Integration and Generation of Unweighted Distributions". [Online]. http://xxx.lanl.gov/abs/arXiv:0709.2085

[81] E. de Doncker et al. Parint: Parallel Integration Package. [Online]. http://www.cs.wmich.edu/parint/

[82] Fukuko Yuasa, Tadashi Ishikawa, Setsuya Kawabata, Denis Perret-Gallix, Kazuhiro Itakura, Yukihiko Hotta, Motoi Okuda, "Hybrid parallel computation of integration in GRACE," in *6th International Workshop on New Computing Techniques in Physics Research (AIHEP99)*. [Online]. http://arxiv.org/abs/hep-ph/0006268

[83] Giuliano Laccetti, Marco Lapegna, Valeria Mele, Diego Romano, Almerico Murli, "A Double Adaptive Algorithm for Multidimensional Integration on Multicore Based HPC Systems," *International Journal of Parallel Programming*, vol. 40, no. 4, pp. 397-409, 2012. [Online]. http://rd.springer.com/article/10.1007%2Fs10766-011-0191-4

[84] F Yuasa, T Ishikawa, N Hamaguchi, T Koike, N Nakasato, "Acceleration of Feynman loop integrals in high-energy physics on many core GPUs," in *Conference on Computational Physics (Kobe)*, 2012. [Online]. http://acpp-coll.in2p3.fr/cgi-bin/twiki.source/pub/ACPP/PresentationsNotes/ccp2012-poster-v2.pdf

[85] J. Kanzaki, "Monte Carlo integration on GPU," *Eur. Phys. J. C* , vol. 71, p. 1559. [Online]. http://arxiv.org/abs/1010.2107

[86] K. Kondo, "Dynamical likelihood method for reconstruction of events with missing momentum. 2: Mass spectra for 2 → ; 2 processes," *J.Phys.Soc.Jap.*, vol. 60, pp. 836–844, 1991.

[87] Madweight. [Online]. https://cp3.irmp.ucl.ac.be/projects/madgraph/wiki/MadWeight

[88] Robert Harrington, "Optimization of Matrix element methods using GPUs," in *Workshop on Future Computing, University of Edinburgh*, 2011. [Online]. http://indico.cern.ch/getFile.py/access?contribId=10&sessionId=3&resId=0&materialId=slides&confId=141309

[89] O. Brandt, "Top Quark mass mesurement at Tevatron," *Il Nuovo Cimiento*, vol. 33, no. C.N.4, pp. 73-80, 2010.

[90] Walter T. Giele, Ciaran Williams John M. Campbell, "Extending the matrix element method to next-to-leading order," FERMILAB-CONF-12-176-T, FERMILAB-CONF-12-176-T. [Online]. http://arxiv.org/pdf/1205.3434.pdf

[91] CDF Collaboration, "Measurements of the Top-quark Mass and the tt-bar Cross Section in the Hadronic tau + Jets Decay Channel at sqrt(s)=1.96 TeV". [Online]. http://arxiv.org/pdf/1208.5720v4.pdf

[92] J. Fujimoto, T. Ishikawa, D. Perret-Gallix, "High precision numerical computations: A case for an HAPPY design," ACPP IRG note: ACPP-N-1 2005. [Online]. http://emc2.in2p3.fr/cgi-bin/twiki.source/pub/ACAT/PresentationsNotes/Highprecisionnumericalcomputatio3.pdf

[93] Mian Lu, Bingsheng He, Qiong Luo, "Supporting Extended Precision on Graphics Processors," in *DaMoN '10 Proceedings of the Sixth International Workshop on Data Management on New Hardware*, pp. 19-26. [Online]. http://www.ntu.edu.sg/home/bshe/precision_damon10.pdf

[94] Sverre Jarp, Alfio Lazzaro, Andrzej Nowak, Liviu Valsan, "Comparison of Software Technologies for Vectorization and Parallelization," CERN openlab, 2012. [Online]. http://openlab.web.cern.ch/sites/openlab.web.cern.ch/files/technical_documents/Evaluation_of_Parallel_Technology.pdf

[95] Xin Dong, Gene Cooperman and John Apostolakis, "Multithreaded Geant4: Semi-Automatic Transformation into Scalable Thread-Parallel Software," College of Computer Science, Northeastern University, Boston and PH/SFT, CERN (Switzerland),. [Online]. http://citeseerx.ist.psu.edu/viewdoc/download?doi=10.1.1.174.8538&rep=rep1&type=pdf

[96] Lynn Garren, Paul Lebrun. StdHep: StdHep provides a common output format for Monte Carlo events. [Online]. http://cepa.fnal.gov/psm/stdhep/

[97] M. Dobbs and J.B. Hansen, "HepMC - a C++ Event Record for Monte Carlo Generators," *Comput. Phys. Commun.*, vol. 134, p. 41, 2001. [Online]. https://savannah.cern.ch/projects/hepmc/

[98] T. Binoth, F. Boudjema, G. Dissertori, A. Lazopoulos, A. Denner, S. Dittmaier, R. Frederix, N. Greiner, S. Hoche, W. Giele, P. Skands, J. Winter, T. Gleisberg, J. Archibald, G. Heinrich, F. Krauss, D. Maitre, M.Huber, J. Huston, N. Kauer, et al., "A proposal for a standard interface between Monte Carlo tools and one-loop programs". [Online]. http://xxx.lanl.gov/abs/1001.1307